# Can the Journal Impact Factor Be Used as a Criterion for the Selection of Junior Researchers? A Large-Scale Empirical Study Based on ResearcherID Data

Lutz Bornmann* & Richard Williams**

*Division for Science and Innovation Studies

Administrative Headquarters of the Max Planck Society

Hofgartenstr. 8,

80539 Munich, Germany.

E-mail: bornmann@gv.mpg.de

**Department of Sociology

810 Flanner Hall

University of Notre Dame

Notre Dame, IN 46556 USA

E-mail: Richard.A.Williams.5@ND.Edu

Web Page: http://www3.nd.edu/~rwilliam/

**Abstract**

Early in researchers' careers, it is difficult to assess how good their work is or how important or influential the scholars will eventually be. Hence, funding agencies, academic departments, and others often use the Journal Impact Factor (JIF) of where the authors have published to assess their work and provide resources and rewards for future work. The use of JIFs in this way has been heavily criticized, however. Using a large data set with many thousands of publication profiles of individual researchers, this study tests the ability of the JIF (in its normalized variant) to identify, at the beginning of their careers, those candidates who will be successful in the long run. Instead of bare JIFs and citation counts, the metrics used here are standardized according to Web of Science subject categories and publication years. The results of the study indicate that the JIF (in its normalized variant) is able to discriminate between researchers who published papers later on with a citation impact above or below average in a field and publication year – not only in the short term, but also in the long term. However, the low to medium effect sizes of the results also indicate that the JIF (in its normalized variant) should not be used as the sole criterion for identifying later success: other criteria, such as the novelty and significance of the specific research, academic distinctions, and the reputation of previous institutions, should also be considered.

**Keywords**

Journal Impact Factor, ResearcherID, Web of Science, Normalized citation scores, Journal metrics

# 1    Introduction

Processes for selecting researchers are prevalent in science. Promising candidates are selected for fellowships, post-doctoral positions, professorships, etc. As a rule, the peer review process is used to separate the wheat from the chaff (Bornmann, 2011). For example, the European Molecular Biology Organization's (EMBO) Long-Term Fellowships support postdoctoral research visits to laboratories worldwide (see http://www.embo.org/funding-awards/fellowships/long-term-fellowships). All applications are evaluated by the EMBO Fellowship Committee, which bases its funding decision on (1) previous scientific achievements, (2) novelty and significance of proposed research, and (3) appropriateness of the host laboratory for the proposed research (see http://www.embo.org/funding-awards/fellowships/long-term-fellowships#selection) (Bornmann, Wallon, & Ledin, 2008). As is common in many other selection processes, bibliometrics is a decisive factor in the EMBO selection process: applicants for a fellowship "must have at least one first (or joint first) author research paper accepted for publication, in press or published in an international peer-reviewed journal at the time the EMBO Long-Term Fellowships application is complete" (see http://www.embo.org/documents/LTF/LTF_Guidelines_for_Applicants.pdf).

In order to assess the importance, quality or impact of publications, many reviewers and administrative staff of funding organizations use the Journal Impact Factor (JIF, Clarivate Analytics, formerly the Intellectual Property & Science business of Thomson Reuters) of the journals in which the applicants have published their papers (Wouters et al., 2015). The JIF is available in the Journal Citation Reports (JCR) and measures the average citations in one year (e.g., 2014) of the journal's papers that were published in the two preceding years (e.g., 2012 and 2013). Since the JIF is easily accessible for many researchers (and beyond), and since evaluated units (e.g., scientists) have, as a rule, published more than one paper in a journal, the use of the JIF for impact measurement is attractive. Thus, JIFs often serve as a proxy for paper-level citation statistics for evaluating professionals. The results of van Dijk, Manor, and Carey (2014) show that the JIF is an important factor in becoming a principle investigator in biomedicine. From the point of view of Elsevier (the provider of the Scopus database), the JIF is such an important journal metric in research evaluation that they introduced the CiteScore which resembles the JIF (https://journalmetrics.scopus.com/, https://www.cwts.nl/blog?article=n-q2y254).

In recent years, the practice of basing funding decisions (mainly) on the JIF has been heavily criticized – also by the inventor of the JIF (Garfield, 2006). The most important reasons given are that (1) the JIF measures citation impact for a very short time period only; and (2) since the JIF is an average value that is based on skewed citation distributions, it cannot represent the citation impact of most of the journal's papers (Seglen, 1992). Recently, the *San Francisco Declaration on Research Assessment* (see http://www.ascb.org/dora) appeared as a statement against the use of the JIF for the evaluation of individual papers and their authors (Garwood, 2013). By November 28, 2016, 12,583 individuals and 916 institutions had signed the declaration. However, according to Hutchins, Yuan, Anderson, and Santangelo (2016) "a groundswell of support for the San Francisco Declaration on Research Assessment … has not yet been sufficient to break this cycle. Continued use of the JIF as an evaluation metric will fail to credit researchers for publishing highly influential work." Reich (2013) reports that publishing in high-impact journals leads to bonuses or salary increases for researchers in some developing countries.

Based on a large data set with many thousands of individual researchers' publication profiles, this study investigates whether the practice of using the JIF in research evaluation processes makes sense or whether the JIF should be eliminated from these processes. To answer these questions, the researchers' publication profiles are separated into a starting block of publication activity at the beginning of their careers (the first five years) and a subsequent block of about ten years as a senior researcher. The study tests whether the ability of researchers to publish in high-impact journals (during the first five years) is related to the citation impact of the papers published after the initial period. In other words, do researchers who started their career by publishing in high-impact journals perform outstanding research later on as measured by field- and time-normalized citation scores of individual publications?

This study follows initiatives like that of Waltman and Traag (2017) who try to link the JIF discussion with sound theoretical and empirical analyses. Their computer simulations point out that the JIF "is a more accurate indicator of the value of an article than the number of citations the article has received".

## 2     Literature overview

Since the current study is intended to investigate the relationship between different metrics for individual researchers, the literature overview refers to studies that examine the relationship of several metrics at the level of individual researchers. Only a small portion of these studies compare the metrics at different points in time (e.g., at the beginning and end of the academic career). Several studies investigating individual researchers' careers deal with the relationship between productivity (proxy of quantity) and citation impact (proxy of quality). Most of these studies demonstrate that there is a strong correlation between quantity and quality (see an early overview in Hemlin, 1996). Researchers who publish frequently seem to write the best papers, and vice versa: "highly cited researchers are also highly productive" (Parker, Allesina, & Lortie, 2013, p. 469). Abramo, D'Angelo, and Costa (2010) were able to show in a large-scale study including 26,000 researchers working in the Italian university system, that "scientists who are more productive in terms of quantity also achieve higher levels of quality in their research products" (p. 139). Also, van den Besselaar and Sandström (2015) report a positive correlation between number of publications and number of highly cited papers for researchers in the Swedish science system. The positive correlation exists not only on the size-dependent level (number of publications and citations), but also on a mix of size-dependent and size-independent levels: number of publications and citations per publication (Diem & Wolter, 2013).

According to the results of Larivière and Costas (2016), the positive "quantity-quality" correlation can be observed especially for biomedical and health sciences, and for social sciences and humanities. Costas, Bordons, van Leeuwen, and van Raan (2009) concretise the positive "quantity-quality" correlation using the publication profiles of 1,064 researchers working as scientific staff at the Consejo Superior de Investigaciones Científicas (CSIC): they found that "researchers in low field-citation-density regions and those whose impact is below world class tend to benefit the most from an increase in number of publications" (p. 750). The positive "quantity-quality" correlation reported in several studies might confirm the cumulative advantage theory of Merton (1968) and the reinforcement theory of Cole and Cole (1973). Both theories claim that current successful researchers (in terms of publications, citations, funds, etc.) are likely to be more successful in the future (in terms of publications, citations, funds, etc.).

Several other empirical studies on researchers' publication profiles have investigated the skewed distribution of publications across the profiles (Abramo, D'Angelo, & Soldatenkova, 2017; Piro, Rørstad, & Aksnes, 2016; Ruiz-Castillo & Costas, 2014). Although it has been observed that the number of papers per researcher has increased, particularly in recent decades (Fanelli & Larivière, 2016), the productivity patterns of researchers are very different. Ioannidis, Boyack, and Klavans (2014) found 15,153,100 publishing researchers in the entire Scopus database. Only less than 1% have published a paper in every year of the period under review (between 1996 and 2011). This core set of researchers accounts for 42% of all papers, and for 87% of papers with more than 1,000 citations. Similar patterns of differences between researchers are also reported in small-scale studies: there are many education science professors with no publications, and only a handful of professors who frequently publish (Diem & Wolter, 2013). Possible reasons for different publication patterns are provided by Amara, Landry, and Halilem (2015): "scholars who publish frequently and are frequently cited differ from those in the low performing profile in many ways: they are full professors, they dedicate more time to their research activities, they receive all their research funding from research councils, and, finally, they are located in top tier universities" (p. 489). The latter result is confirmed by Yang, Rousseau, Huang, and Yan (2015): most top scientists work at top organizations.

In the final paragraphs of this literature overview, some studies that predict success for individual researchers using bibliometric data are presented. The acceptance of the recent study by Sinatra, Wang, Deville, Song, and Barabási (2016) on publication profiles of researchers from multiple disciplines in *Science* demonstrates the great general relevance and topicality of this topic. They studied the publication records of 2,887 physicists over a period of at least 20 years. A second data set consists of 24,630 Google Scholar career profiles from multiple disciplines. Sinatra et al. (2016) found that the growth of productivity is more pronounced for high-impact researchers and is modest for low-impact researchers. Similar patterns were observed for the growth of citation impact in both groups. In a follow-up study to the introduction of the *h* index by Hirsch (2005), Hirsch (2007) tested whether the *h* index has predictive power and can be used to select subsequently successful researchers. This study is also based on physicists' publication profiles as divided into two parts: a beginning period and a follow-up period of 12 years. His results indicate that "the *h* index and the total number of citations are better than the number of papers and the mean citations per paper to predict future achievement, with achievement defined by either the indicator itself or the total citation count $N_c$" (p. 19196).

Laurance, Useche, Laurance, and Bradshaw (2013) identified nearly 200 researchers working in the biological and environmental sciences. They were interested in the question of whether the academic success of these researchers – measured in terms of the number of peer-reviewed papers following their PhD – can be predicted by factors in effect around the time they completed their PhD. Their findings suggest the following conclusions: long-term publication success is closely related to early publication success. Furthermore, the long-term scientific impact of the publications can be enhanced by publishing in high-impact journals and by frequent collaboration with other researchers. The authors conclude that their findings "highlight a crucial role for early training and mentorship for aspiring academics … the best way to promote the long-term success of one's graduate students is to assist them in publishing early and establishing this as a key performance indicator for both students and their graduate supervisors" (p. 821).

This study continues the line of research which shows that long-term publication success is related to early publication success by focusing on the use of journal metrics for assessing early careers. Early in researchers' careers, it is difficult to assess how good their work is or how important or influential the scholars will eventually be. Hence, funding agencies, academic departments, and others often use journal metrics of where the authors have published to assess their work and provide resources and rewards for future work. The use of JIFs in this way has been heavily criticized, however. JIFs (or journal metrics) alone can easily overlook highly influential and innovative work. In this paper, we address these concerns by examining how early standardized JIF scores are related to the individual citation impact of scholars years later. In contrast to many other studies published to date, the current study is based solely on field- and time-normalized impact scores and a large sample of researcher profiles.

# 3 Methods

## 3.1 Data set used

The bibliometric data used in this study are from an in-house database developed and maintained by the Max Planck Digital Library (MPDL, Munich) and derived from the Science Citation Index Expanded (SCI-E), Social Sciences Citation Index (SSCI), and Arts and Humanities Citation Index (AHCI) prepared by Clarivate Analytics. The bibliometric data were matched with the ResearcherID (RID, www.researcherid.com) of 272,921 researchers; the RIDs are available in the in-house database of the Competence Centre for Bibliometrics (www.bibliometrie.info). The researchers published a total of 6,495,715 articles (this study is restricted to the document type "article" in order to have comparable units of analysis). Thus, the researchers published, on average, 24 articles between 1948 (the earliest paper in the data set) and 2012. Papers published later than 2012 were excluded from the study in order to ensure a citation window of at least five years for every paper (Glänzel & Schöpflin, 1995). Citations were counted until 2016 in the MPDL in-house database.

RID provides a possible solution to the author ambiguity problem within the scientific community. The problem of polysemy means, in this context, that multiple authors are merged in a single identifier; the problem of synonymy entails multiple identifiers being available for a single author (Boyack, Klavans, Sorensen, & Ioannidis, 2013). Each researcher is assigned a unique identifier in order to manage his or her publication list. The difference between this and similar services provided by Elsevier within the Scopus database is that Elsevier automatically manages the publication profiles of researchers (authors), with the profiles being able to be manually revised. With RID, researchers themselves take the initiative, create a profile, and manage their publication lists. Although it cannot be taken for granted that the publication lists on RID are error-free, these lists will probably be more reliable than the automatically generated lists (by Elsevier).

For this study, not all available profiles in our data set of 272,921 researchers are used. A selected set of three cohorts with comparable researchers have been separated out: The first cohort consists of 3,976 researchers who published their first paper in 1998 and at least one paper in 2012. One can expect that these researchers published more or less continuously over 15 years. The publication periods of the researchers have been separated into two parts, allowing the first five years of their careers (as junior scientists) to be compared with the remaining ten years (as senior scientists). The second (n=4,517) and third (n=4,687) researcher cohorts published their first papers in 1999 and 2000, respectively. Since these researchers also published at least one paper in 2012, their careers could be divided into a first part of five years (as junior scientists) and a second part of nine years and eight years (as senior scientists), respectively. In section 4, the results for the first cohort of researchers with an RID are presented in detail. The results for the second (1999-2012) and third (2000-2012) cohorts are used to contrast the results for the first cohort. We want to make sure that there aren't quirks across time or in the data that affects one cohort differently than it does others.

What are the reasons for these criteria for the selection of the cohorts? A key problem is that we do not know what scholars dropped out of academia. They may not have gotten tenure; or, they may have gone off into the private sector where publishing was not expected or even allowed. To address these concerns, we require that every member of our sample has to have both early publications and at least one publication in 2012. This greatly increases the likelihood that our sample includes researchers who remained active across time, while still allowing for the possibility that the quality of their work may have varied. This does allow for potential biases in our sample (e.g. prolific scholars who didn't have a paper in 2012 are excluded; while scholars who started out fast but then "burned out" by 2012 are also missed). But, we think it does have the advantage of letting us examine how quality of work varied for active scholars across time.

Since the researchers included in this study are from different disciplines (which are, unfortunately, unknown), only field-normalized citation impact scores are used. The impact of the journals in which the researchers published within the first five years of their careers was measured, not with the raw JIF (which is not field- and time normalized), but with a percentile-based measure that goes back to Pudovkin and Garfield (2004). They proposed to rank the journals in each WoS subject category by their JIF and to identify the $x$% most frequently cited journals (Wouters et al., 2015). This percentile-based measure is used in the SCImago Institutions Ranking (SIR, see http://www.scimagoir.com/methodology.php), and is known as the Q1 indicator (see also Bornmann, Stefaner, de Moya Anegón, & Mutz, 2014). Q1 is the proportion of an institution's papers that have been published in the 25% most frequently cited journals of the corresponding subject categories (Miguel, Chinchilla-Rodriguez, & de Moya-Anegón, 2011). According to Liu, Hu, and Gu (2016), one can expect that approximately 45% of the publications are published in first-quartile journals. Bornmann and Marx (2014) propose to use the Q1 indicator for individual researchers, too. It can be applied for the comparison of researchers from different fields (which is not possible with the JIF, see Leydesdorff, Wouters, & Bornmann, 2016) and shows whether researchers are able to publish in reputable journals (measured in terms of citation impact).

The number of papers is used as a second indicator of academic success at the beginning of the career to contrast the Q1 results. It is common practice in academia to assess (young) researchers by the number of their (peer reviewed) papers. The results of the studies presented in section 2 reveal that high citation impact can be expected for researchers who publish frequently. Although the publication numbers of researchers are not field-normalized, they are used for comparison with the normalized Q1 indicator. It is still not usual in bibliometrics to normalize publication counts. The results of D'Angelo and Abramo (2015) show that the average intensity of publication is extremely variable also within the same discipline. So, it is not clear on which field-level the normalization should be undertaken. Furthermore, the results of Koski, Sandström, and Sandström (2016) point out that it is necessary to create advanced productivity indicators. For example, they recommend estimating the zero-class of a truncated paper frequency distribution. Thus, there are still so many open questions in normalizing productivity metrics that we abstain from normalization in this study.

In order to measure the field- and time-normalized impact of individual publications in the later careers of the researchers, we use the standard approach in bibliometrics: the normalized citation score (NCS). Here, the citation counts of a focal paper are divided by an expected value. The expected value is the mean citation rate of those papers that have been published in the same WoS subject category and publication year as the focal paper. Although improved field- and time-normalized indicators have been introduced in recent years (e.g., percentiles or citing side indicators) (Bornmann & Marx, 2015), the MNCS is widely used in databases (e.g., Scopus) and university rankings (e.g., the Times Higher Education Ranking). To go from the individual publication level to the researcher level, NCS values were aggregated into the size-independent mean normalized citation score (MNCS, Waltman, van Eck, van Leeuwen, Visser, & van Raan, 2011) and the size-dependent total normalized citation score (TNCS, Abramo & D'Angelo, 2016a, 2016b; Waltman, 2016). The MNCS was calculated as the mean NCS value of the publications of a researcher; the TNCS was calculated as the sum of the NCS values. Although the two measures tend to be highly correlated, some scientometricians prefer size-independent indicators while others prefer size-dependent. Including both will show how robust the findings are using different popular and widely-used approaches.

### 3.2 Statistical methods

This study uses inferential statistics for comparing the citation performance of several groups of researchers (Williams & Bornmann, 2016). According to Claveau (2016) the general argument for using inferential statistics in citation analysis is "that these observations are realizations of an underlying data generating process constitutive of the research unit. The goal is to learn properties of the data generating process. The set of observations to which we have access, although they are all the *actual* realizations of the process, do not constitute the set of all *possible* realizations. In consequence, we face the standard situation of having to infer from an accessible set of observations – what is normally called the sample – to a larger, inaccessible one – the population. Inferential statistics are thus pertinent" (p. 1233).

In this study, the Q1 indicator is basically the independent variable (JIF-type measure for the first five years) and the NCS is the dependent variable – the measure of the researchers' paper level citation scores. The Q1 indicator is used to assign the researchers to $k$ groups. This study explores whether the NCS from $k$ groups (where $k > 2$) are the same or not. The analysis of variance (ANOVA) can be used to detect any overall difference between the $k$ groups that is statistically significant. The ANOVA is a method for the variance analysis because it separates the variance components into those due to mean differences and those due to random influences (Riffenburgh, 2012). There are three assumptions required for calculating the ANOVA in this study: (1) The publication data are independent of each other. (2) The citation impact distribution of papers for each researcher group is normal. (3) The standard deviation of the citation impact data is the same for all groups. Although the assumptions are violated in this study, the ANOVA is still used: according to Riffenburgh (2012), the ANOVA "is fairly robust against these assumptions" (p. 265), especially in those cases where the sample size of the study is high. In order to confirm the results of the ANOVA, the Kruskal-Wallis rank test (KW test) has been additionally calculated as the non-parametric alternative (Acock, 2016).

Following the ANOVA, the effect size eta squared ($\eta^2$) is calculated as a measure of the practical significance of the results (Acock, 2016). Eta squared is the sum of squares for a factor (here: members of different groups) divided by the total sum of squares. The effect size shows how much of the citation impact variation in the sample of researchers is explained by the factor. According to Cohen (1988), a value of $\eta^2 = 0.01$ is a small effect, $\eta^2 = 0.06$ a medium effect, and $\eta^2 = 0.14$ a large effect. Measures of effect sizes are especially important in situations where the case numbers in a study are very high (Kline, 2004) – as is the case in this study. In these situations, the results of statistical tests may be significant although the effects (e.g., mean citation impact differences between two groups) are small.

# 4  Results

Table 1 shows the MNCS and TNCS for researchers who published (1) a different proportion of papers in first-quartile journals (Q1 indicator) and (2) a different number of papers (NP) between 1998 and 2002. The MNCS and TNCS is based on papers published between 2003 and 2012. In other words, the success of the publication activities in terms of field- and time-normalized citations is shown for researchers who publish with different levels of success at the beginning of their careers (measured in terms of Q1 and NP).

The results of the ANOVA in Table 1 show that the MNCS differences between the groups of researchers publishing different numbers of papers in first-quartile journals is statistically significant, $F(3,3,972) = 63.31$, $p = .000$.[1] The KW test confirms the statistically significant results of the ANOVA, $\chi^2(3) = 355.64$, $p = .000$. With $\eta^2 = .05$, the effect size for the difference between the groups of researchers is small to medium (Cohen, 1988). The results for the TNCS are similar. Both indicators separate clearly between the four groups. For example, the results indicate that the MNCS and TNCS are higher for researchers who published between 75% and 100% in first-quartile journals than those for researchers who published (1) between 50% and 75%, (2) between 25% and 50%, and (3) up to 25% of their papers in first-quartile journals.

---

[1] We used the style of the American Psychological Association (2010) for the reporting of statistics (e.g., *M* is the abbreviation of the Mean).

Table 1. Mean normalized citation scores (MNCS) and total normalized citation scores (TNCS) for researchers who published (1) a different proportion of papers in first-quartile journals (Q1 indicator) and (2) a different number of papers (NP) between 1998 and 2002. The MNCS is based on papers published between 2003 and 2012. The correlation coefficients show the spearman rank correlation between the MNCS and TNCS, respectively, and the proportion of papers in first-quartile journals and the number of papers, respectively.

| Q1 indicator | MNCS | | TNCS | | Number of researchers |
|---|---|---|---|---|---|
| | Mean | Standard deviation | Mean | Standard deviation | |
| First quartile (up to 25%) | 1.02 | 0.79 | 23.06 | 34.15 | 1,082 |
| Second quartile (between 25% and 50%) | 1.24 | 0.83 | 36.99 | 49.19 | 568 |
| Third quartile (between 50% and 75%) | 1.42 | 1.07 | 43.67 | 56.09 | 948 |
| Fourth quartile (between 75% and 100%) | 1.72 | 1.78 | 50.31 | 79.36 | 1,378 |
| Total | 1.39 | 1.31 | 39.41 | 60.93 | 3,976 |
| **NP** | | | | | |
| First quartile (lowest number) | 1.24 | 1.19 | 23.24 | 39.01 | 1,474 |
| Second quartile | 1.44 | 1.73 | 32.16 | 40.70 | 918 |
| Third quartile | 1.44 | 1.01 | 43.26 | 60.84 | 617 |
| Fourth quartile (highest number) | 1.53 | 1.17 | 68.49 | 87.86 | 967 |
| Total | 1.39 | 1.31 | 39.41 | 60.93 | 3,976 |

Notes. MNCS, Q1 indicator: $F(3, 3,972) = 63.31$, $p = .000$, $\eta^2 = .05$ [.03, .06], $\chi^2(3) = 355.64$, $p = .000$, $r_s=.30$
MNCS, NP: $F(3, 3,972) = 11.29$, $p = .000$, $\eta^2 = .008$ [.003, .014], $\chi^2(3) = 97.13$, $p = .000$, $r_s=.16$
TNCS, Q1 indicator: $F(3, 3,972) = 43.89$, $p = .000$, $\eta^2 = .03$ [.02, .04], $\chi^2(3) = 315.60$, $p = .000$, $r_s=.27$
TNCS, NP: $F(3, 3,972) = 123.62$, $p = .000$, $\eta^2 = .09$ [.07, .10], $\chi^2(3) = 667.80$, $p = .000$, $r_s=.41$

For the purpose of contrasting the results based on the proportion of papers published in first-quartile journals, Table 1 shows the results for four groups of researchers that have been categorized using their different proportions of papers. The expectation in this study is that the use of the Q1 indicator in selecting candidates has greater predictive power than the number of papers, since the later citation impact of papers is also dependent on the ability of researchers to publish high-quality papers. The probability of publishing in high-impact journals is greater for high-quality papers than for low-quality ones. In this study, the number of articles within the first five years of their careers is computed for every researcher. Then, the researchers are categorized into quartiles based on their publication counts: The researchers in the first quartile ($n = 1,474$) have the lowest number of articles ($M = 2.03$). The researchers in the further quartiles with higher numbers have published the following mean and maximum number of papers: second quartile ($n = 918$): $M = 4.45$; third quartile ($n = 617$): $M = 6.38$; and fourth quartile ($n = 967$): $M = 11.91$.

As the results of the comparison between the four researcher groups that published different numbers of papers show, the MNCS difference is also statistically significant, $F(3, 3,972) = 11.29$, $p = .000$, $\eta^2 = .008$ [.003, .014], $\chi^2(3) = 97.13$, $p = .000$. This result is in agreement with the results based on the group assignment using Q1. However, the $\eta^2$ as a measure of practical significance reveals that the relationship is very small. For the TNCS, the results are different from that for the MNCS. Since the TNCS is a size-dependent variable, it ($r_s=.41$) correlates significantly higher with NP than the MNCS ($r_s=.16$).

Taken as a whole, the results indicate that Q1 is better able than the number of articles to predict the ability of researchers to publish their papers with a relatively high mean citation impact in their later careers. However, if the later success of the researchers is measured in terms of the sum of the normalized scores (TNCS), the NP outperforms Q1.

Table 2. Mean normalized citation scores (MNCS) and total normalized citation scores (TNCS) for researchers who published (1) a different proportion of papers in first-quartile journals (Q1 indicator) and (2) a different number of papers (NP) between 1999 and 2003. The MNCS is based on papers published between 2004 and 2012. The correlation coefficients show the spearman rank correlation between the MNCS and TNCS, respectively, and the proportion of papers in first-quartile journals and the number of papers, respectively.

| **Q1 indicator** | MNCS | | TNCS | | Number of researchers |
| --- | --- | --- | --- | --- | --- |
| | Mean | Standard deviation | Mean | Standard deviation | |
| First quartile (up to 25%) | 1.01 | 0.78 | 20.31 | 41.62 | 1,175 |
| Second quartile (between 25% and 50%) | 1.20 | 0.82 | 33.87 | 52.70 | 684 |
| Third quartile (between 50% and 75%) | 1.43 | 1.01 | 39.52 | 69.43 | 1,105 |
| Fourth quartile (between 75% and 100%) | 1.83 | 1.80 | 53.33 | 103.68 | 1,553 |
| Total | 1.42 | 1.32 | 38.41 | 76.87 | 4,517 |
| **NP** | | | | | |
| First quartile (lowest number) | 1.26 | 1.30 | 21.01 | 42.43 | 1,698 |
| Second quartile | 1.37 | 1.02 | 28.06 | 48.90 | 566 |
| Third quartile | 1.52 | 1.60 | 37.27 | 56.67 | 1,138 |
| Fourth quartile (highest number) | 1.59 | 1.12 | 71.35 | 122.94 | 1,115 |
| Total | 1.42 | 1.32 | 38.41 | 76.87 | 4,517 |

Notes. MNCS, Q1 indicator: $F(3, 4,513) = 101.31$, $p = .000$, $\eta^2 = .06$ [.05, .08], $\chi^2(3) = 539.47$, $p = .000$, $r_s=.35$

MNCS, NP: $F(3, 4,513) = 16.78$, $p = .000$, $\eta^2 = .01$ [.005, .017], $\chi^2(3) = 137.55$, $p = .000$, $r_s=.17$

TNCS, Q1 indicator: $F(3, 4,513) = 43.26$, $p = .000$, $\eta^2 = .03$ [.02, .04], $\chi^2(3) = 469.68$, $p = .000$, $r_s=.31$

TNCS, NP: $F(3, 4,513) = 107.94$, $p = .000$, $\eta^2 = .07$ [.05, .08], $\chi^2(3) = 786.75$, $p = .000$, $r_s=.42$

Table 3. Mean normalized citation scores (MNCS) and total normalized citation scores (TNCS) for researchers who published (1) a different proportion of papers in first-quartile journals (Q1 indicator) and (2) a different number of papers (NP) between 2000 and 2004. The MNCS is based on papers published between 2005 and 2012. The correlation coefficients show the spearman rank correlation between the MNCS and TNCS, respectively, and the proportion of papers in first-quartile journals and the number of papers, respectively.

| **Q1 indicator** | MNCS | | TNCS | | Number of researchers |
|---|---|---|---|---|---|
| | Mean | Standard deviation | Mean | Standard deviation | |
| First quartile (up to 25%) | 1.00 | 0.77 | 16.89 | 21.55 | 1,175 |
| Second quartile (between 25% and 50%) | 1.27 | 0.87 | 28.86 | 35.80 | 737 |
| Third quartile (between 50% and 75%) | 1.44 | 1.15 | 36.19 | 56.81 | 1,150 |
| Fourth quartile (between 75% and 100%) | 1.73 | 1.71 | 41.82 | 80.28 | 1,625 |
| Total | 1.40 | 1.30 | 32.15 | 58.64 | 4,687 |
| **NP** | | | | | |
| First quartile (lowest number) | 1.31 | 1.54 | 19.73 | 38.98 | 1,652 |
| Second quartile | 1.38 | 1.04 | 25.57 | 37.20 | 1,053 |
| Third quartile | 1.46 | 1.27 | 33.21 | 42.93 | 1,009 |
| Fourth quartile (highest number) | 1.53 | 1.12 | 59.26 | 97.69 | 973 |
| Total | 1.40 | 1.30 | 32.15 | 58.64 | 4,687 |

Notes. MNCS, Q1 indicator: $F(3, 4,683) = 76.78$, $p = .000$, $\eta^2 = .05$ [.04, .06], $\chi^2(3) = 461.55$, $p = .000$, $r_s = .31$

MNCS, NP: $F(3, 4,683) = 7.01$, $p = .000$, $\eta^2 = .01$ [.001, .009], $\chi^2(3) = 94.09$, $p = .000$, $r_s = .14$

TNCS, Q1 indicator: $F(3, 4,683) = 45.08$, $p = .000$, $\eta^2 = .03$ [.02, .04], $\chi^2(3) = 397.39$, $p = .000$, $r_s = .27$

TNCS, NP: $F(3, 4,683) = 105.11$, $p = .000$, $\eta^2 = .06$ [.05, .08], $\chi^2(3) = 753.70$, $p = .000$, $r_s = .40$

Table 2 and Table 3 show the MNCS and TNCS for researchers who published (1) a different proportion of papers in first-quartile journals (Q1 indicator) and (2) a different number of papers (NP) between 1999 and 2003 (see Table 2) and between 2000 and 2004 (see Table 3), respectively. The NCS is based on papers published between 2004 and 2012 (see Table 2) and between 2005 and 2012 (see Table 3). These additional analyses using two further cohorts of researchers were undertaken in order to see whether the results in Table 1 (which is based on the periods 1998-2002 and 2003-2012) are stable or not. As the results in both tables show, the initial results were confirmed. The more frequently researchers published in high-impact journals in the corresponding subject categories at the beginning of their careers, the higher the MNCS of their papers in subsequent years. The comparison of the groupings based on Q1 and NP reveals that the (time- and field-normalized) JIF is better able than the NP to identify promising candidates (see especially Cohen's *d* values). Although the TNCS is also different between the four researcher groups which have been formed on the base of the Q1 indicator, the TNCS differs especially for researchers with varying numbers of papers.

In the second part of the statistical analysis, it was investigated whether the citation impact of the papers published by the different groups decreases or increases across the publication years after the initial five-year period. An increase in citation impact would accord with the cumulative advantage theory (see above). The results of the second part of the statistical analyses are presented in Figure 1, Figure 2, and Figure 3. The results in Figure 1 are based on two periods (in accordance with the results in Table 1): the initial five-year period between 1998 and 2002, and the subsequent years between 2003 and 2012. These results are contrasted with the results for two further cohorts in Figure 2 and Figure 3 (in accordance with the results in Table 2 and Table 3). The figures show dot plots for the annual normalized citation impact scores (MNCS and TNCS) of the papers published by the four groups (which were built on the base of the Q1 indicator and the NP, respectively). To calculate the mean impact per year in the figures, first the annual mean impact (MNCS) and annual total impact (TNCS), respectively, across the papers of a researcher in one year was calculated. Then these impact scores were arithmetically averaged for the annual dot plots in the figures.

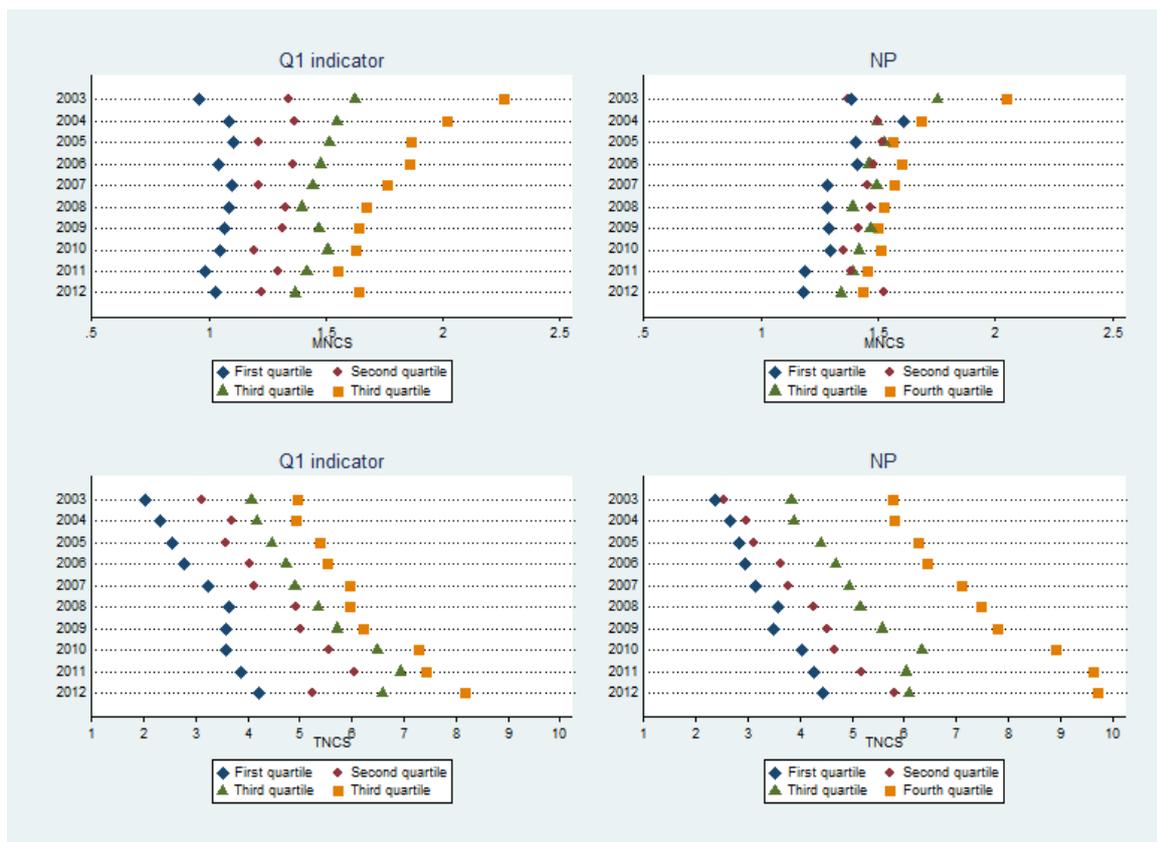

Figure 1. Mean normalized citation scores (MNCS) and total normalized citation scores (TNCS) for researchers who published (1) a different proportion of papers in first-quartile journals (Q1 indicator) and (2) a different number of papers (NP) between 1998 and 2002. The mean MNCS and TNCS values are shown for the papers published between 2003 and 2012.

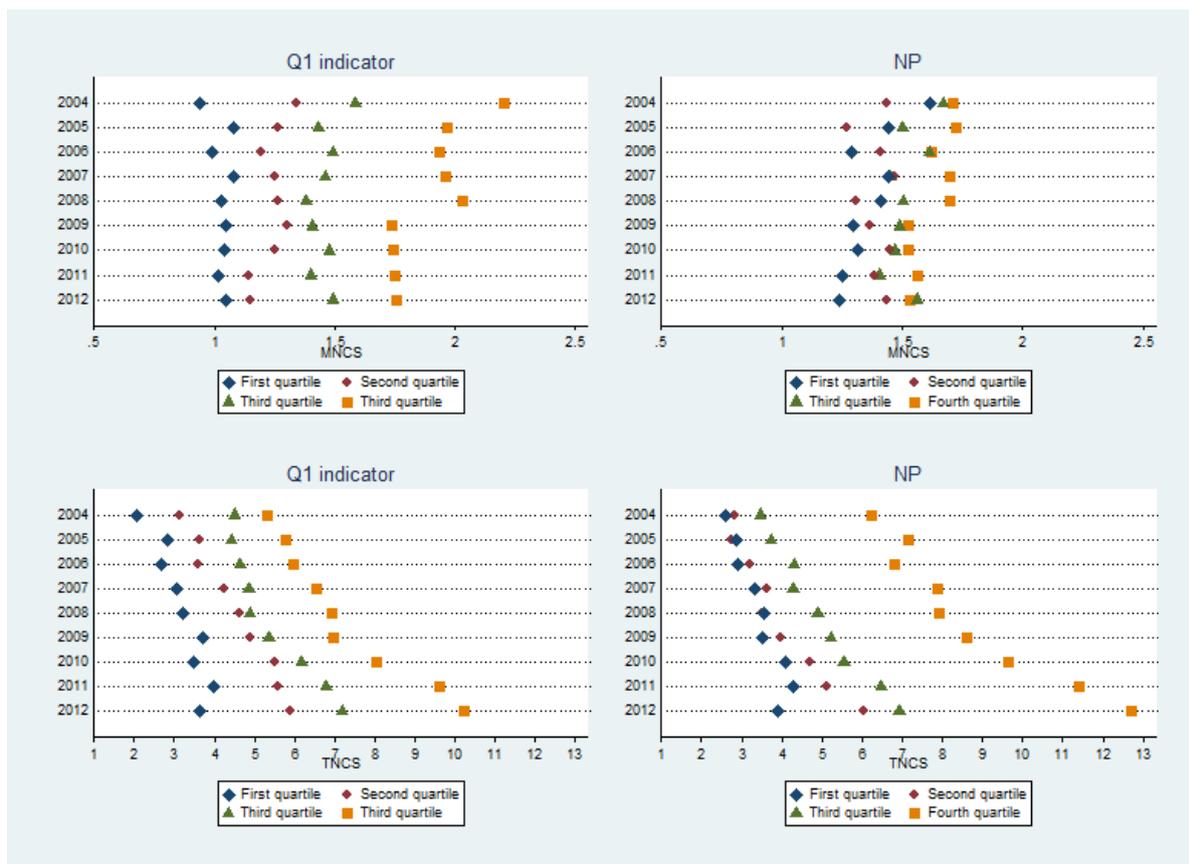

Figure 2. Mean normalized citation scores (MNCS) and total normalized citation scores (TNCS) for researchers who published (1) a different proportion of papers in first-quartile journals (Q1 indicator) and (2) a different number of papers (NP) between 1999 and 2003. The mean MNCS and TNCS values are shown for the papers published between 2004 and 2012.

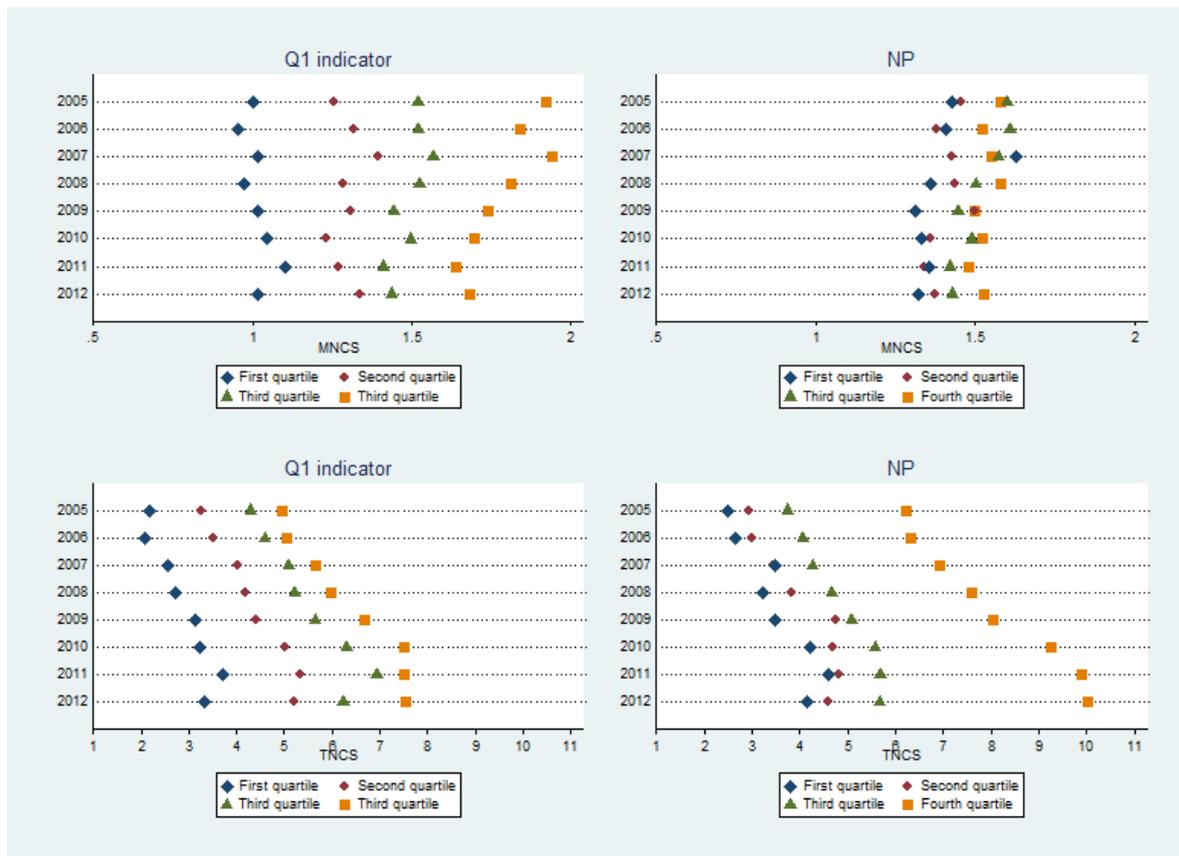

Figure 3. Mean normalized citation scores (MNCS) and total normalized citation scores (TNCS) for researchers who published (1) a different proportion of papers in first-quartile journals (Q1 indicator) and (2) a different number of papers (NP) between 2000 and 2004. The mean MNCS and TNCS values are shown for the papers published between 2005 and 2012.

The results in Figure 1, Figure 2, and Figure 3 look very similar. They confirm the results from Table 1, Table 2, and Table 3. It can be expected, especially from people who publish frequently in high-impact journals at the beginning of their careers, that they will publish papers with an above-average citation impact later on. Furthermore, the JIF – in its normalized variant – seems to differentiate more or less successfully between promising and uninteresting candidates not only in the short term, but also in the long term. However, there are differences visible between the results based on the Q1 indicator and NP. The researcher groups which have been formed on the base of NP differ significantly in terms of the TNCS, but not in terms of the MNCS.

If we take a look at the annual impact scores in the figures in the long term, we see different distributions for the MNCS and TNCS. For the MNCS, there is no increase of citation impact visible in the figures of the top-group. Thus, the results do not seem to accord with the cumulative advantage theory. Instead, the results seem to suggest a tendency towards the center across the years. This tendency could have performance-based reasons: The initially weak researchers (in terms of Q1) become stronger and the initially strong researchers become weaker. Another explanation for the results, however, is related to the technique of impact measurement. The citation window for the impact measurements in Figure 1, Figure 2, and Figure 3 is not fixed, but variable from publication year until 2016. Suppose that the publications from the initially strong researchers (in terms of Q1) need a long time period in general to show their high value for other researchers. Then, one can expect a decreasing annual average impact for their publications, because the citation window becomes smaller. We cannot test this possible explanation with our data, because we do not have the normalized impact values for fixed time periods. However, the alternative explanation would better accord to the results which are based on the TNCS: there is an increase visible for the distributions. Here, the results definitely agree to the cumulative advantage theory.

## 5  Discussion

This study is part of research efforts in scientometrics which analyze the correlation between metrics for single researchers. Whereas most of the studies in this area focus on the relationship between quality (measured by citations) and quantity (measured by paper numbers) (see the overview in section 2), this study was intended to investigate the relationship between the JIF of journals in which researchers have published in their early careers and the citation impact of the papers which have been published later on.

Using three cohorts of researchers who have published across a long period of time, we tested the ability of the JIF (in its normalized variant) to identify, at the beginning of their careers, those candidates who are successful in the long run. Compared with previous studies, this study is based on a broad data set and uses field- and time-normalized data. Instead of bare JIFs and citation counts, the metrics used here are standardized according to WoS subject categories and publication years. The results of the study indicate that the JIF (in its normalized variant as Q1) is able to discriminate between researchers who published papers later on with a citation impact above or below average in a field and publication year – not only in the short term, but also in the long term. Our study shows that early success in publishing in high-impact journals is related to later success with individual-level citations. At the same time, the relationship is far from deterministic. Further, there are styles of scholarship, e.g. book-writing, where JIFs are probably not a good measure of quality or potential. Just like university admissions committees should not rely solely on standardized test scores, university departments and granting agencies should not rely solely on early JIFs (in their normalized variants) when rewarding work and allocating resources.

If we discuss the results against the backdrop of theoretical approaches that deal with academic success in the long run (see section 2), the results seem to confirm the "sacred spark" theory: "there are substantial, predetermined differences among scientists in their ability and motivation to do creative scientific research" (Allison & Stewart, 1974, p. 596). Researchers who can publish several papers in high-impact journals at the beginning of their careers seem to be essentially able to do creative research. This ability seems to persist in subsequent careers, which finds expression in high citation scores for the papers. The researchers are probably "motivated by an inner drive to do science and by a sheer love of the work" (Cole & Cole, 1973, p. 62). They seem to have the capacity "to work hard and persist in the pursuit of long-range goals" (Fox, 1983, p. 287).

However, the results of this study can also be interpreted otherwise. It is possible that researchers do well later on because they benefited from early publications in reputable journals. In other words, the findings could also be explained by the fact that reputable journals are part of the selection mechanisms in science (e.g., for hiring and funding at top universities). If this is the case, then the results do not suggest that junior researchers "should" be selected on the basis of journal metrics, but that they "have been" selected on their base. This reasoning is part of the cumulative advantage theory of Merton (1968),

In the discussion of the JIF for the use of research evaluation purposes, one should bear in mind that metrics might serve as incentives that influence the motivational forces of researchers (see here Bornmann, 2012; Rijcke, Wouters, Rushforth, Franssen, & Hammarfelt, 2015): "The JIF has the potential to deeply change the motivational forces of scientists. In a world where the JIF determines the value of a publication, successful goal-directed behavior requires knowledge about the JIF and the ability to make use of this knowledge when potential publications are within reach. However, it remains unknown how scientists have incorporated the JIF into their reward circuitry and adapted their behavior in order to match the affordances for 'survival' in academia" (Paulus, Rademacher, Schäfer, Müller-Pinzler, & Krach, 2015). The results of the project "The impact of indicators: How evaluation shapes biomedical knowledge production" (de Rijcke & Rushforth, 2015; Rushforth & de Rijcke, 2015) by Alex Rushforth and Sarah de Rijcke from the Center for Science and Technology Studies (CWTS) at the Leiden University show that this indicator should not be dismissed "as mere idle 'publication talk,' or as floating in some external 'cultural' realm separated from the 'serious business' of knowledge making. Likewise statements of discontent tend to implicate the entire field of biomedicine as captured by the JIF" (Rushforth & de Rijcke, 2015, p. 136).

We need more scientometric research on the effects of the use of the JIF (and similar metrics) on knowledge production in different fields. Does knowledge production suffer from the use of journal metrics for research evaluation purposes? Are promising candidates neglected in peer review processes if these processes rely mainly on journal metrics? Future studies should also consider the possibility that knowledge production in the fields profits from the focus of peers and decision makers on the JIF (and similar metrics). The focus of researchers on high-impact journals that, as a rule, use the peer review process could lead to an increase in publications reporting reliable and important results. Thus, knowledge production in the fields might profit from evaluation practices that focus on journal metrics. If NP is included in these future studies (as we did it here for comparison with Q1), researchers should try to consider (additionally) field-normalized productivity measures. Koski et al. (2016) published ways to calculate these measures.

In this study, we included only researchers staying in science (in all likelihood). The bias of exclusively dealing with researchers that have had a longer career affects the results of this study. Perhaps researchers that quit science (or stopped publishing) did not continue in science exactly because they published in low-impact journals. But since we do not know what their later performance would have been (in terms of bibliometrics), we do not know whether they have been "rightly" not stayed in science. Perhaps, they would have had some brilliant publications, but we will never know. Of course it may also be that researchers with excellence performance went on to work in industry because they are the most appealing to the R&D departments. The point is that this bias can obfuscate the results. Therefore, future studies should not only consider researchers staying in science, but also those leaving science.

A somewhat related point is the use of the RID in this study to identify the publications of single researchers. Researchers that actively maintain their publication list could perhaps have a higher performance overall – in terms of productivity and citation impact. On the other hand, it could also be that researchers only maintain their RID because they still need it: those who already have tenure no longer maintain it. The exact effect on the results is not clear, but future studies should (additionally) use other databases (besides RID) to identify publications of single researchers.

# Acknowledgements

The bibliometric data used in this paper are from an in-house database developed and maintained by the Max Planck Digital Library (MPDL, Munich) and derived from the Science Citation Index Expanded (SCI-E), Social Sciences Citation Index (SSCI), and Arts and Humanities Citation Index (AHCI) prepared by Clarivate Analytics, formerly the IP & Science business of Thomson Reuters.